\documentclass[12pt]{article}

\usepackage{amsmath}
\usepackage{amsfonts}
\usepackage{amsthm}

\usepackage{url}

\newcommand{\reals}{\mathbf{R}}
\newcommand{\sphere}{\mathbf{S}^1}
\newcommand{\torus}{\mathbf{T}}

\newcommand{\U}{\mathcal{U}}

\renewcommand{\d}{\partial}
\newcommand{\rmd}{\mathrm{d}}
\newcommand{\rme}{\mathrm{e}}
\newcommand{\tr}{\operatorname{tr}}
\newcommand{\cov}[1]{
  \nabla_{\mskip-1.5\thinmuskip #1}}%
\newcommand{\Dplus}{D_{\!+}}
\newcommand{\Dminus}{D_{\!-}}

\newcommand{\abs}[1]{\lvert#1\rvert}
\newcommand{\norm}[1]{\lVert{#1}\rVert}

\newcommand{\phieo}{\phi_{e_0}}
\newcommand{\phiei}{\phi_{e_1}}

\let\underscore=\_
\renewcommand{\_}{\mskip-\thinmuskip_}
\let\asciicircum=\^
\renewcommand{\^}{\mskip-0.5\thinmuskip^}

\newtheorem{theorem}{Theorem}[section]
\newtheorem{lemma}[theorem]{Lemma}
\newtheorem{corollary}[theorem]{Corollary}

\newcommand{\updatebox}{\fbox{\small\sffamily\bfseries Update}}
\newcommand{\update}{\marginpar{\updatebox}}
  {\begin{sffamily}\small\update\begin{itemize}}%
  {\end{itemize}\end{sffamily}}

\begin{document}

\title{Shock Waves in Plane Symmetric Spacetimes}
\author{Alan D.\ Rendall%
        \thanks{E-mail address: rendall@aei.mpg.de} \\
        Max-Planck-Institut f\"ur Gravitationsphysik\\
        Albert-Einstein-Institut\\
        Am M\"uhlenberg~1, D-14476 Golm, Germany
        \\[2ex]
        Fredrik St{\aa}hl%
        \thanks{E-mail address: Fredrik.Stahl@miun.se} \\
        Dept. for Engineering, Physics and Mathematics\\
        Mid Sweden University, S-83125 \"Ostersund, Sweden
       }
\date{}

\maketitle

\begin{abstract}
  We consider Einstein's equations coupled to the Euler equations in
  plane symmetry, with compact spatial slices and constant mean
  curvature time. We show that for a wide variety of equations of
  state and a large class of initial data, classical solutions break
  down in finite time. The key mathematical result is a new theorem on
  the breakdown of solutions of systems of balance laws. We also show
  that an extension of the solution is possible if the spatial
  derivatives of the energy density and the velocity are bounded,
  indicating that the breakdown is really due to the formation of
  shock waves.
\end{abstract}


\section{Introduction}
\label{sec:intro}

A question of central interest in general relativity is that of the
long-time behaviour of self-gravitating matter. Mathematically the
starting point is to get suitable existence and uniqueness theorems for
the Einstein equations coupled to the equations of motion of the
matter. Since the resulting system of partial differential equations
is difficult to handle it makes sense to begin with solutions of high
symmetry. One of the most popular matter models in applications is the
perfect fluid described by the relativistic Euler equations. It is to
be expected that, as in classical hydrodynamics, a major difficulty in
studying the long-time behaviour of solutions of these equations is
the formation of shocks from smooth initial data. If the solution is
to be continued beyond these it is necessary to leave the realm of
classical solutions of the equations. It should be noted that until 
very recently most of the theorems on solutions of the Euler equations
involving shocks were a one-dimensional context (plane symmetric solutions).
This may change following the recent work of Christodoulou on the formation
of shocks in special relativity without symmetry assumptions. 

One way to avoid the difficulties involved with fluids is to consider
instead collisionless matter described by the Vlasov equation. In that
case quite a lot of mathematical results are available for the coupled
Einstein-matter equations
\cite{Rendall:existence-review,Andreasson:Einstein-Vlasov-review}. If,
on the other hand, we face the problems associated with the fluid
description, as we do in this paper, several natural questions arise.
First, are there global existence theorems for weak solutions of the
special relativistic Euler equations? Some positive answers have been
given in \cite{Smoller-Temple:relativistic-Euler},
\cite{Chen:conserv-SR} and \cite{Frid-Perepelitsa:rel-gas-dynamics}.
Second, do these results extend to the case of a self-gravitating
fluid? Theorems have been proved under certain assumptions by
\cite{Groah-Temple:EE-fluid-Glimm} and by
\cite{Barnes-etal:Glimm-Gowdy}. Third, can it be proved that classical
solutions of the Einstein-Euler equations break down in finite time?
It is the third question which is addressed in the following.

The symmetry assumed in this paper is plane symmetry where the
solutions are invariant under the action of the full isometry group of
the Euclidean plane. In particular this reduces the full problem to a
problem in one time and one space dimension. It is assumed that the
position space is compact. This circumvents the need for boundary
conditions and is the analogue of periodic boundary conditions in the
non-relativistic case. The notion of 'finite time' breakdown is subtle
in general relativity. In that theory there is a free choice of time
variable and if it has been proved that a solution breaks down after a
finite amount of a particular time coordinate it is necessary to think
carefully about what this means geometrically. This is what is
relevant for physics. In this paper we use a constant mean curvature
(CMC) time coordinate. This means that the value of $t$ at some point
of spacetime is equal to the mean curvature of the unique compact CMC
hypersurface passing through that point.

The main result of this paper (Theorem~\ref{th:blowup-theorem}) shows
that plane symmetric classical solutions of the Einstein-Euler
equations exhibit finite-time breakdown for a wide variety of
equations of state. The key new analytical result used to prove this
is a theorem on the breakdown of solutions of systems of balance laws
which is of interest in its own right. This is combined with general
estimates for the Einstein equations to give the final result.

In the last section we put the result into context, considering
several issues. The Euler equations are compared with other matter
models. The assumptions on the equation of state are compared with
cases known in the literature, and the relations to existing results
on weak solutions are discussed.


\section{The Einstein equations}
\label{sec:Einstein}

Let $(M,g)$ be a spacetime, where the manifold is assumed to be
$M=I\times\torus^3$, $I$ is a real interval and
$\torus^3=\sphere\times\sphere\times\sphere$ is the three-torus. We
require that the metric $g$ and the matter fields are invariant under
the action of the Euclidean group $E_2$ on the universal cover, and
that the spacetime has an $E_2$-invariant Cauchy surface of constant
mean curvature (CMC).  As was shown in \cite{Rendall:crushing-sings},
there is a local in time $3+1$ decomposition of $(M,g)$ where each
spatial slice has constant mean curvature. We can introduce spatial
coordinates $x$, $y$ and $z$ on each slice, with ranges $[0,2\pi]$ and
period $2\pi$, and a CMC time coordinate $t=\tr{k}$ where $k$ is the
second fundamental form of the slices.

The metric can be expressed as
\begin{equation}
  \label{eq:Einstein-ds2}
  \rmd s^2 =
  -\alpha^2 \rmd t^2
  + A^2 \bigl[
          (\rmd{x}+\beta\rmd{t})^2
          + a^2(\rmd{y}^2+\rmd{z}^2)
        \bigr],
\end{equation}
where $\alpha$, $A$ and $\beta$ depend on $t$ and $x$ and $a$ depends
on $t$ only. Here $\alpha$ is the lapse and $\beta\delta^\mu\_1$ is
the shift. The coordinates can be chosen such that
$\int_0^{2\pi}\!\beta\,\rmd{x}=0$.

It is convenient to introduce the orthonormal frame
\begin{equation}
\label{eq:Einstein-e}
\begin{split}
  e_0 &= \alpha^{-1}\d_t - \alpha^{-1}\beta\d_x, \\
  e_1 &= A^{-1}\d_x, \\
  e_2 &= (aA)^{-1}\d_y, \\
  e_3 &= (aA)^{-1}\d_z.
\end{split}
\end{equation}
We will use indices $a,b,\dots$ to denote spatial coordinate
components and $i,j,\dots$ to denote spatial frame components.

The frame components of the second fundamental form may be written
\begin{equation}
  \label{eq:Einstein-kij}
  k_{ij} = - \tfrac12 (K-t) \delta_{ij}
           + \tfrac12 (3K-t) \delta_{1i} \delta_{1j},
\end{equation}
where $K$ is a function of $t$ and $x$. We also introduce the notation
$\rho=T^{00}$, $j=T^{10}$ and $S^{ik}=T^{ik}$ for frame components
of the energy momentum tensor $T^{\mu\nu}$. The Einstein equations can
then be written
\begin{subequations}
  \label{eq:Einstein}
  \begin{gather}
    \label{eq:Einstein-momentum}
    K' + (3K - t) A^{-1} A'
    = 8 \pi A j, \\
    \label{eq:Einstein-Hamiltonian}
    (\sqrt{\!A})''
    = - \tfrac18 A^{5/2}
           \bigl[
             K^2 + \tfrac12 (K-t)^2 - t^2 + 16\pi \rho
           \bigr], \\
    \label{eq:Einstein-lapse}
    \alpha'' + A^{-1}A'\alpha'
    = \alpha A^2 
         \bigl[
           K^2 + \tfrac12 (K-t)^2 + 4\pi(\rho + \tr{S})
         \bigr]
       - A^2, \\
    \label{eq:Einstein-dadt}
    \d_t{a} = a \bigl[ - \beta' + \tfrac12 \alpha (3K-t) \bigr], \\
    \label{eq:Einstein-dAdt}
    \d_t{A} = - \alpha KA + (A\beta)', \\
    \label{eq:Einstein-dKdt}
    \begin{split}
      \d_t{K} =&\ \beta K' - A^{-2}\alpha'' + A^{-3} A'\alpha' \\
               &+ \alpha \bigl[
                           - 2A^{-3}A'' + 2A^{-4}(A')^2 + Kt
                           - 4\pi ( 2S_{11} - \tr{S} + \rho )
                         \bigr].
    \end{split}
\end{gather}
\end{subequations}
Here a prime denotes differentiation by $x$.
Equation~\eqref{eq:Einstein-momentum} and
\eqref{eq:Einstein-Hamiltonian} are the momentum and Hamiltonian
constraints, respectively. The lapse equation
\eqref{eq:Einstein-lapse} comes from the constant mean curvature
condition, while \eqref{eq:Einstein-dadt} and \eqref{eq:Einstein-dAdt}
are consequences of the choice of spatial coordinate conditions. The
last equation \eqref{eq:Einstein-dKdt} is the only independent
Einstein evolution equation in this case.

Using~\eqref{eq:Einstein} it is possible to show that many of the
fundamental quantities are bounded. This was done
in~\cite{Rendall:crushing-sings} for plane, hyperbolic and spherical
symmetry, when the matter satisfies the dominant energy and the
non-negative pressures conditions. The non-negative pressures
condition was subsequently relaxed to the strong energy condition
in~\cite{Rendall:CMC-2dim-symmetry} (note that plane symmetry is a
special case of local $U(1)\times{U}(1)$ symmetry). Also, a bound for
$\alpha^{-1}$ in terms of the previously obtained geometric bounds was
found in~\cite{Rendall:exist-nonexist-CMC}.

\begin{theorem}[See \cite{Rendall:crushing-sings},
  \cite{Rendall:CMC-2dim-symmetry} and
  \cite{Rendall:exist-nonexist-CMC}]%
  \label{th:Einstein-bounds}
  Let a solution of the Einstein equations with plane symmetry be
  given and suppose that when coordinates are chosen which cast the
  metric into the form~\eqref{eq:Einstein-ds2} with constant mean
  curvature time slices the time coordinate takes all values in the
  finite interval $[t_0,t_1)$ with $t_1<0$.  Suppose further that the
  dominant and strong energy conditions hold.  Then the following
  quantities are bounded on the interval $[t_0,t_1)$:
  \begin{equation}
    \label{eq:Einstein-bounded-vars}
    \alpha, \alpha^{-1}, \alpha',
    A, A^{-1}, A', \d_t{A}, 
    a, a^{-1}, \d_t{a}, 
    K, \beta, \beta'.
  \end{equation}
  The bounds involve only
  \begin{equation}
    \label{eq:Einstein-bounds}
    t_0, t_1, \norm{a}_0, \norm{a^{-1}}_0, \norm{A}_0, \norm{A^{-1}}_0,
  \end{equation}
  where $\norm{\cdot}_0$ is the supremum norm on the initial surface
  $t=t_0$.
\end{theorem}


\section{The Euler equations}
\label{sec:Euler}

In this section we will rewrite the matter equations for a perfect
fluid in plane symmetry as a system of balance laws.

The basic matter variables are the pressure $p$, the energy density
$\mu$ and the unit 4-velocity $U^\mu$ of the fluid. Because of the
plane symmetry the frame components of $U$ can be written as
\begin{equation}
  \label{eq:Euler-u}
  U^0 = (1-u^2)^{-1/2},
  \qquad
  U^1 = u(1-u^2)^{-1/2}
  \qquad\text{and}\qquad
  U^2 = U^3 = 0,
\end{equation}
where $u\in(-1,1)$.
The energy momentum tensor for a perfect fluid is
\begin{equation}
  \label{eq:Euler-T}
  T^{\mu\nu}=(\mu+p)U^{\mu}U^{\nu}+{p}g^{\mu\nu},
\end{equation}
which implies 
\begin{subequations}
  \label{eq:Euler-T-frame}
  \begin{align}
    \rho = T^{00} &= \frac{(\mu+p)}{1-u^2} - p, \\
    j = T^{01} &= \frac{(\mu+p)u}{1-u^2}, \\
    S^{11} = T^{11} &= \frac{(\mu+p)u^2}{1-u^2} + p, \\
    S^{22} = S^{33} &= T^{22} = T^{33} = p.
\end{align}
\end{subequations}
It will be convenient to introduce two new variables
\begin{subequations}
\label{eq:Euler-wphidef}
\begin{align}
  \label{eq:Euler-wdef}
  w &= \Bigl( \frac{\rmd{p}}{\rmd\mu} \Bigr)^{1/2}, \\
  \label{eq:Euler-phidef}
  \varphi &= \int_{\mu_-}^{\mu}
                   \Bigl( \frac{\rmd{p}}{\rmd{m}} \Bigr)^{1/2}
                   \frac{\rmd{m}}{m+p},
\end{align}
\end{subequations}
where $m$ is a dummy integration variable and the constant $\mu_-$ is
arbitrary.

The matter equations are given by the vanishing of the divergence
$\cov{\nu}T^{\nu\sigma}$ of the energy momentum tensor. The spatial
frame components $\cov{\nu}T^{\nu2}$ and $\cov{\nu}T^{\nu3}$ vanish
identically. Expressing the remaining two components in terms of
$\varphi$, $w$ and $u$ gives
\begin{subequations}
  \label{eq:Euler1}
  \begin{multline}
    \label{eq:Euler1a}
    w \bigl[ 2 u e_0(u) + (1+u^2) e_1(u) \bigr] \\
    + (1-u^2) \bigl[ (1+u^2w^2) e_0(\varphi)
                     + u (1+w^2) e_1(\varphi) \bigr] \\
    - w (1-u^2) \bigl[ K u^2 + t - 2uA^{-1}e_1(A)
                       - 2u\alpha^{-1}e_1(\alpha) \bigr]
    = 0,
  \end{multline}
  \begin{multline}
    \label{eq:Euler1b}
    w \bigl[ (1+u^2) e_0(u) + 2 u e_1(u) \bigr] \\
    + (1-u^2) \bigl[ u(1+w^2) e_0(\varphi)
                     + (u^2+w^2) e_1(\varphi) \bigr] \\
    - w (1-u^2) \bigl[ (K+t)u - 2u^2 A^{-1}e_1(A)
                       - (1+u^2)\alpha^{-1}e_1(\alpha) \bigr]
    = 0.
  \end{multline}
\end{subequations}
Adding and subtracting \eqref{eq:Euler1a} and \eqref{eq:Euler1b}
we get
\begin{subequations}
  \label{eq:Euler2}
  \begin{multline}
    E_0 =
    w\bigl[ e_0(u) + e_1(u) \bigr]
    + (1-u)\bigl[ (1+uw^2)e_0(\varphi) + (u+w^2) e_1(\varphi) \bigr] \\
    - w(1-u)\bigl[ Ku + t - 2u A^{-1}e_1(A)
                   - (1+u) \alpha^{-1}e_1(\alpha) \bigr]
    = 0,
  \end{multline}
  \begin{multline}
    E_1 =
    w\bigl[ e_0(u) - e_1(u) \bigr]
    - (1+u)\bigl[ (1-uw^2)e_0(\varphi) + (u-w^2) e_1(\varphi) \bigr] \\
    - w(1+u)\bigl[ Ku - t + 2u A^{-1}e_1(A)
                   - (1-u) \alpha^{-1}e_1(\alpha) \bigr]
    = 0.
  \end{multline}
\end{subequations} 
The linear combinations $(1+u)(1\pm w)E_0+(1-u)(1\mp w)E_1$ give
\begin{subequations}
  \label{eq:Euler3}
  \begin{align}
    \Dplus u + (1-u^2) \Dplus \varphi
    &= (1-u^2) \biggl[
                 \frac{K u + t w - 2 u w A^{-1} e_1(A)}{1+uw}
                 - \alpha^{-1} e_1(\alpha)
               \biggr], \\
    \Dminus u - (1-u^2) \Dminus \varphi
    &= (1-u^2) \biggl[
                 \frac{K u - t w + 2 u w A^{-1} e_1(A)}{1-uw}
                 - \alpha^{-1} e_1(\alpha)
               \biggr],
  \end{align}
\end{subequations}
with differentiation operators
\begin{equation}
  \Dplus = e_0 + \frac{u + w}{1 + uw} e_1
  \qquad\text{and}\qquad
  \Dminus = e_0 + \frac{u - w}{1 - uw} e_1.
\end{equation}

Next, we introduce the variables
\begin{equation}
  \label{eq:Euler-Riemann}
  r = \varphi + \frac12 \ln\frac{1+u}{1-u}
  \qquad\text{and}\qquad
  s = \varphi - \frac12 \ln\frac{1+u}{1-u}.
\end{equation}
These are analogues of the Riemann invariants found by Taub
\cite{Taub:Rankine-Hugoniot}. In contrast with the special
relativistic case, $r$ and $s$ are not invariant. In particular, it
follows from~\eqref{eq:Euler3} that instead of a system of
conservation laws, $r$ and $s$ satisfy a system of balance laws:
\begin{subequations}
  \label{eq:Euler-balance}
  \begin{align}
    \Dplus r
    &= \frac{t w + K u - 2 u w A^{-1} e_1(A)}{1+uw}
       - \alpha^{-1} e_1(\alpha), \\
    \Dminus s
    &= \frac{t w - K u - 2 u w A^{-1} e_1(A)}{1-uw}
       + \alpha^{-1} e_1(\alpha).
\end{align}
\end{subequations}
Here $u$ and $w$ should be regarded as functions of $r$ and $s$
given by \eqref{eq:Euler-Riemann} and \eqref{eq:Euler-wphidef}.

We will now make some assumptions on the equation of state to ensure
that the maps between the different matter variables introduced in
this section are well behaved.
\begin{lemma}
  \label{la:Euler-smooth}
  Suppose that the equation of state is given as a smooth function $p$
  of $\mu$ such that
  \begin{enumerate}
  \item\label{cond:Euler-SEC}
    the strong energy condition ($\mu+p\ge0$ and $\mu+3p\ge0$) holds,
  \item\label{cond:Euler-DEC}
    the dominant energy condition ($-\mu\le p \le\mu$) holds,
  \item\label{cond:Euler-dpdmu}
    $0<\frac{\rmd{p}}{\rmd\mu}<1$ and
  \item\label{cond:Euler-d2pdmu2}
    $\frac{\rmd^2{p}}{\rmd\mu^2}\ge0$.
  \end{enumerate}
  Then the maps
  \begin{gather}
    \label{eq:Euler-smooth-uphi}
    (r,s) \mapsto (u,\varphi), \quad \reals^2 \to (-1,1)\times\reals, \\
    \label{eq:Euler-smooth-mu}
    \mu \mapsto \varphi, \quad (0,\infty) \to (\varphi_-,\infty),
                         \quad -\infty\le\varphi_-<0
  \end{gather}
  are smooth bijections,
  \begin{equation}
    \label{eq:Euler-smooth-rhojS}
    (\mu,u) \mapsto (\rho,j,S)
  \end{equation}
  is smooth and 1--1 on $(0,\infty)\times(-1,1)$,
  \begin{equation}
    \label{eq:Euler-smooth-w}
    \mu \mapsto w, \quad (0,\infty) \to (0,1) \\    
  \end{equation}
  is smooth and increasing, and
  \begin{equation}
    \label{eq:Euler-smooth-dwdphi}
    \frac{\rmd{w}}{\rmd\varphi}\ge0.
  \end{equation}
\end{lemma}

\begin{proof}
  From \eqref{eq:Euler-Riemann},
  \begin{equation}
    \label{eq:Euler-rstophiu}
    \varphi = \frac{r+s}2
    \qquad\text{and}\qquad
    u = \tanh \frac{r-s}2,
  \end{equation}
  which establishes \eqref{eq:Euler-smooth-uphi} without any
  restrictions on the equation of state.  Applying
  condition~\ref{cond:Euler-SEC} and \ref{cond:Euler-dpdmu}
  to~\eqref{eq:Euler-phidef} shows that $\varphi$ is a smooth
  strictly increasing function of $\mu$ and thus 1--1. Moreover, from
  condition~\ref{cond:Euler-d2pdmu2} and \ref{cond:Euler-DEC} it
  follows that
  \begin{equation}\label{eq:Euler-phi+infty}
    \lim_{\mu\to\infty}\varphi=\infty.
  \end{equation}
  Thus $\varphi$ is a smooth bijection from $(0,\infty)$ to
  $(\varphi_-,\infty)$, where
  $\varphi_-=\lim_{\mu\to0^+}\varphi\ge-\infty$.

  That $\rho$, $j$ and $S$ are smooth functions of $\mu$ and $u$ is
  evident from~\eqref{eq:Euler-T-frame}. The map is 1--1 since $p$ is
  1--1 because of condition~\ref{cond:Euler-dpdmu}.  Finally,
  \eqref{eq:Euler-smooth-w} follows directly from
  condition~\ref{cond:Euler-dpdmu} and \ref{cond:Euler-d2pdmu2}, and
  \begin{equation}
    \frac{\rmd{w}}{\rmd\varphi}
    = \frac{\rmd{w}}{\rmd\mu} \Big/ \frac{\rmd\varphi}{\rmd\mu}
    = \frac12 (\mu + p) \Bigl(\frac{\rmd{p}}{\rmd\mu}\Bigr)^{-1}
                         \frac{\rmd^2{p}}{\rmd\mu^2}
    \ge 0
  \end{equation}
  because of condition~\ref{cond:Euler-SEC}, \ref{cond:Euler-dpdmu}
  and \ref{cond:Euler-d2pdmu2}.
\end{proof}


\section{Balance laws}
\label{sec:balance}

In this section we will obtain a blowup result for the system of
balance laws
\begin{subequations}
\label{eq:balance-laws}
\begin{align}
  \label{eq:balance-laws-r}
  \Dplus r  = e_0(r) + \kappa(r,s)  e_1(r) &= f(t,x,r,s), \\
  \label{eq:balance-laws-s}
  \Dminus s = e_0(s) + \lambda(r,s) e_1(s) &= g(t,x,r,s),
\end{align}
\end{subequations}
where $\{e_0,e_1\}$ is a pseudo-orthonormal frame with respect to a
Lorentzian metric on $[t_0,t_1)\times\sphere$ for some real interval
$[t_0,t_1)$. The operators $\Dplus=e_0+{\kappa}e_1$ and
$\Dminus=e_0+{\lambda}e_1$ are called characteristic derivatives and
we denote the corresponding integral curves (or
\emph{characteristics}) by $\varrho_x$ and $\sigma_x$. The suffix
shows where the curves intersect the surface $t=t_0$, i.e.,
$\varrho_x(0)=(t_0,x)$ and $\sigma_x(0)=(t_0,x)$.  We also denote the
commutation coefficients by $c^k\_{ij}$ so that $[e_i,e_j] =
c^k\_{ij}e_k$.

Let $\U_\delta$ be an open subset of $\reals^2$, to be further
specified below.  We assume that at least one of $\kappa$ and
$\lambda$ is genuinely nonlinear on $\U_\delta$, i.e.,
$\d\kappa/\d{r}\neq0$ or $\d\lambda/\d{s}\neq0$. Without loss of
generality we may assume that $\d\kappa/\d{r}>0$. We also assume that
$\kappa-\lambda\ne0$ on $\U_\delta$ so that the system is strictly
hyperbolic.

Let the initial values of $r$ and $s$ be $r_0(x)=r(t_0,x)$ and
$s_0(x)=s(t_0,x)$. We will show that for certain choices of $r_0$ and
$s_0$, $r$ and $s$ are bounded while $\d_x{r}$ or $\d_x{s}$ blow up in
finite time, indicating the presence of a shock wave.  There are
similar results for systems of conservation laws (see, e.g.,
\cite{Dafermos:conserv-laws}), but we cannot hope for the same
generality here. For more specific results, a better control of the
source terms $f$ and $g$ is needed. For example, if the sources are
superlinear in $r$ and $s$ we can expect blowup of $r$ and/or $s$
themselves~\cite{Jenssen-Sinestrari:blowup-scalar-balance-law}. On the
other hand, global existence of smooth solutions have been shown under
certain conditions~\cite{Yong:global-exist-balance-laws}. We will
leave these considerations aside and aim for a more general but less
sharp result.


\subsection{Bounds on $r$ and $s$}
\label{sec:balance-bounds}

First we need to specify the set $\U_\delta$. The initial data
$(r_0,s_0)$ is a smooth map $\sphere\to\reals^2$. The image is a plane
curve with convex hull $\U_0$. For any $\delta>0$ we define
\begin{equation}
  \label{eq:balance-UOmegadef}
  \U_\delta = \{P\in\reals^2;\; d(P,\U_0)<\delta\}
  \qquad\text{and}\qquad
  \Omega_\delta = [t_0,t_1)\times\sphere\times\U_\delta,
\end{equation}
where $d$ is the Euclidean distance in $\reals^2$. Note that
$\U_\delta$ is convex as well, as follows easily from the triangle
inequality.

To be able to perform integrations along lines of constant $r$ in a
well defined way we need to introduce the following construction.
Since $\sphere$ is compact, there are points $(r_-,s_-)$ and
$(r_+,s_+)$ on the initial data curve in $\reals^2$ where $r_0$
attains its minimum and maximum, respectively. These are also global
extremal points of $r$ in $\U_0$ since $\U_0$ is convex. By
construction, $(r_--\delta,s_-)$ and $(r_++\delta,s_+)$ are extremal
points of $r$ in the closure of $\U_\delta$, and we let $s=\gamma(r)$
be the straight line between those two points. We can now state the
following lemma. Here $\norm\cdot$ is the supremum norm on $\sphere$.

\begin{lemma}
  \label{la:balance-gamma}
  The line from $(r,s)$ to $(r,\gamma(r))$ is contained in $\U_\delta$
  for all $(r,s)\in\U_\delta$. Moreover, $\abs{\gamma'} \le
  \delta^{-1}\norm{s_0}$ and $\abs{s-\gamma(r)}<2\norm{s_0}+\delta$
  on $\U_\delta$.
\end{lemma}
\begin{proof}
  Since $r_--\delta<r_++\delta$, the line $s=\gamma(r)$ intersects the
  balls with radius $\delta$ around $(r_-,s_-)$ and $(r_+,s_+)$. It
  follows from the convexity of $\U_\delta$ that $s=\gamma(r)$ is
  contained in $\U_\delta$ except for its endpoints $(r_--\delta,s_-)$
  and $(r_++\delta,s_+)$. The first statement then follows directly
  from the convexity of $\U_\delta$.

  The second statement is just the rough estimate
  \begin{equation}
    \abs{\gamma'} = \frac{\abs{s_+-s_-}}{r_+-r_-+2\delta}
                \le \frac{\norm{s_0}}{\delta},
  \end{equation}
  and the third follows from the fact that
  $\abs{\gamma(r)}\le\norm{s_0}$ and $\abs{s}<\norm{s_0}+\delta$ on
  $\U_\delta$.
\end{proof}

Let
\begin{equation}
  \label{eq:balance-FEdef}
  F = \sup_{\Omega_\delta} \{ \abs{f}, \abs{g} \}
  \qquad\text{and}\qquad
  E = \inf_{\Omega_\delta} \{ e_0\^0 + {\kappa}e_1\^0,
                        e_0\^0 + {\lambda}e_1\^0 \}.
\end{equation}
The following lemma provides a crude estimate of $r$ and $s$.
\begin{lemma}
  \label{la:balance-rsbound}
  If $F$ is finite and $E>0$, any smooth solution $(r,s)$ of
  \eqref{eq:balance-laws} with initial data $(r_0,s_0)$ remains within
  $\U_\delta$ for $t<\min\{t_0 + 2^{-1/2}\delta EF^{-1}, t_1\}$.
\end{lemma}

\begin{proof}
  From \eqref{eq:balance-laws} we have that $\abs{r-r_0} \le F\xi$
  along a characteristic $\varrho_x$ as long as $t<t_1$, where $\xi$
  is the parameter along $\varrho_x$, chosen such that
  $\Dplus=\d/\d\xi$ and $\varrho_x(t_0)=(t_0,x)$. By the definition of
  $\Dplus$, $\rmd{t}/\rmd\xi = e_0\^0+{\kappa}e_1\^0 \ge E$, so $\xi
  \le E^{-1}(t-t_0)$ and $\abs{r-r_0} \le FE^{-1}(t-t_0)$. We can
  obtain similar estimates for $s$ along the characteristics
  $\sigma_x$. Thus the distance between $(r,s)$ and $(r_0,s_0)$ is at
  most $\sqrt2\,FE^{-1}(t-t_0)$, and the conclusion follows.
\end{proof}

As we saw in Lemma~\ref{la:Euler-smooth},
$\mu\mapsto\varphi\colon(0,\infty)\to(\varphi_-,\infty)$ is smooth and
1--1. It is quite possible that $\varphi_-$ is finite, i.e., that
$\mu=0$ for finite $r$ and $s$. This is indeed the case for a
relativistic polytropic perfect fluid, for example.  This complication
can be circumvented by a further restriction on $\delta$.

\begin{lemma}
  \label{la:balance-varphibound}
  Let $\varphi_0 = \frac12 \inf_{x\in\sphere} \{r_0(x)+s_0(x)\}$.  If
  \begin{equation}
    \label{eq:balance-deltabound}
    0 < \delta < \varphi_0 - \varphi_-
  \end{equation}
  then $\varphi>\varphi_-$ on $\U_\delta$.
\end{lemma}

\begin{proof}
  If $\mu=0$ at some point on the initial surface, then
  $\varphi=(r_0+s_0)/2=\varphi_-$ there. Thus the right hand side
  of~\eqref{eq:balance-deltabound} vanishes and the statement is void.
  On the other hand, if $\mu>0$ for all $x\in\sphere$ at $t=0$, the
  right hand side of~\eqref{eq:balance-deltabound} is a positive
  number because $\sphere$ is compact. The conclusion follows from the
  fact that if $(r,s)\in\U_\delta$ then
  \begin{equation}
    \varphi - \varphi_-
    = \frac12 (r + s) - \varphi_-
    > \varphi_0 - \varphi_- - \delta
    > 0.
  \end{equation}
\end{proof}


\subsection{The blowup equation}
\label{sec:balance-blowup-eq}

For a homogeneous system of two conservation laws, it is possible to
show that shock waves will form for suitable initial data by
introducing a new unknown which is a rescaling of the spatial
derivative of $r$. The scale factor can be chosen such that the new
unknown satisfies a differential equation which does not involve
derivatives of $s$.  We follow a similar route, but as we will see
below it is a bit harder to decouple the equations when source terms
are present.

We are interested in the spatial derivative $e_1(r)$. First note that
\begin{equation}
  \label{eq:balance-ecommut}
  [e_1,e_0]
  = c^0\_{10}e_0 + c^1\_{10}e_1
  = c^0\_{10}\Dplus + (c^1\_{10} - \kappa c^0\_{10}) e_1
\end{equation}
from the definition of $\Dplus$. It follows that
\begin{equation}
  \label{eq:balance-Dpluse1}
  \Dplus e_1
  = (e_1 - c^0\_{10}) \Dplus
    - \bigl(
        \kappa_r e_1(r) - \kappa_s e_1(s)
        + c^1\_{10} - \kappa c^0\_{10}
      \bigr) e_1.
\end{equation}
We adopt the convention of denoting partial derivatives by subscripts,
and we also write $f_{e_1}=e_1\^0f_t+e_1\^1f_x$ for the partial frame
derivative of $f$ with respect to $e_1$, i.e., $e_1(f)$ with $r$ and
$s$ are regarded as constants.

Applying the operator $\Dplus{e_1}$ to $r$ and
using~\eqref{eq:balance-laws-r} gives
\begin{multline}
  \label{eq:balance-Dpluse1r}
  \Dplus e_1 (r)
  = - \kappa_r e_1(r)^2 - \kappa_s e_1(s) e_1(r)
    - (c^1\_{10} - \kappa c^0\_{10}) e_1(r) \\
    + f_{e_1} + f_r e_1(r) + f_s e_1(s) - c^0\_{10} f.
\end{multline}
In particular, there is a term quadratic in $e_1(r)$ which might cause
blowup of $e_1(r)$. The problem is that there is a term involving both
$e_1(r)$ and $e_1(s)$. As in the homogeneous case, the mixed term can
be eliminated by introducing a function $h(r,s)$ by
\begin{equation}
  \label{eq:balance-hdef}
  h(r,s) = \int_{\gamma(r)}^s 
             (\kappa-\lambda)^{-1} \frac{\d\kappa}{\d{s}}
           \,\rmd{s},
\end{equation}
where $\gamma$ is defined in Lemma~\ref{la:balance-gamma}.  Now let
$R=-{\rme^h}e_1(r)$.  Differentiating $R$ and simplifying gives
\begin{equation}
  \label{eq:balance-R}
  \Dplus{R} 
  = a_2 R^2 + a_1 R + a_0
  - \rme^h f_s e_1(s),
\end{equation}
where
\begin{subequations}
\label{eq:balance-a}
\begin{align}
  a_2 &= \rme^{-h} \kappa_r, \\
  a_1 &= f_r + f h_r + (\kappa-\lambda)^{-1} g \kappa_s
       + c^1\_{01} - \kappa c^0\_{01}, \\
  a_0 &= - \rme^h ( f c^0\_{01} + f_{e_1} ).
\end{align}
\end{subequations}

We would like to use \eqref{eq:balance-R} to show that $R$ has to blow
up for appropriately chosen initial data. But there is still a
problematic term involving $e_1(s)$ which cannot be estimated using
bounds on $r$ and $s$ alone. This term is not present in the
homogeneous case since then $f_s$ vanish. It is of course possible to
define a quantity similar to $R$ based on $e_1(s)$ instead of
$e_1(r)$. The result is a system of two coupled equations, but then we
cannot use ODE techniques directly.

It is, however, possible to remove the $e_1(s)$ term by the
following procedure. By construction, 
\begin{equation}\label{eq:balance-Dplus->Dminus}
  \Dplus{s} = \Dminus{s} + (\kappa-\lambda)e_1(s)
            = g + (\kappa-\lambda)e_1(s),
\end{equation}
where we have used \eqref{eq:balance-laws-s}.  Since
$\kappa-\lambda\ne0$, we can introduce a function
\begin{equation}
  \label{eq:balance-phidef}
  \phi(t,x,r,s) = - \int_{\gamma(r)}^s
                      \rme^h (\kappa-\lambda)^{-1} f_s
                    \,\rmd{s}.
\end{equation}
By the chain rule, \eqref{eq:balance-Dplus->Dminus} and
\eqref{eq:balance-laws-r},
\begin{equation}
  \label{eq:balance-Dplus-phi}
  \Dplus{\phi}
  = \phieo + \kappa\phiei
  + \phi_s (\kappa-\lambda) e_1(s)
  + \phi_s g + \phi_r f,
\end{equation}
where $\phieo$ and $\phiei$ are the frame partial derivatives of
$\phi$, regarding $r$ and $s$ as constants. Applying
\eqref{eq:balance-phidef} to \eqref{eq:balance-Dplus-phi} gives
\begin{equation}
  \Dplus{\phi} = \phieo + \kappa\phiei
               - \rme^h f_s e_1(s) + \phi_s g + \phi_r f.
\end{equation}
Now \eqref{eq:balance-R} may be written
\begin{equation}
  \label{eq:balance-R-phi}
  \Dplus(R-\phi) = A_2 (R-\phi)^2 + A_1 (R-\phi) + A_0,
\end{equation}
where the coefficients are
\begin{subequations}\label{eq:balance-Adef}
\begin{align}
  A_2 &= \rme^{-h} \kappa_r, \\
  A_1 &= f_r + f h_r + (\kappa-\lambda)^{-1} g \kappa_s
       + c^1\_{01} - \kappa c^0\_{01} + 2 \rme^{-h} \kappa_r \phi, \\
  A_0 &= \rme^{-h} \kappa_r \phi^2
       + \bigl(
           f_r + f h_r + (\kappa-\lambda)^{-1} g \kappa_s
           + c^1\_{01} - \kappa c^0\_{01}
         \bigr) \phi \\
      &\phantom{=\ }
      - \rme^h \bigl( 
                  f c^0\_{01} + f_{e_1} - f_s (\kappa-\lambda)^{-1} g
                \bigr)
      - \phi_r f - \phieo - \kappa\phiei.
\end{align}
\end{subequations}
Note that~\eqref{eq:balance-R-phi} is a first order ODE in $R-\phi$
whose coefficients can be estimated in terms of $r$ and $s$.


\subsection{Blowup of $R$}
\label{sec:balance-blowup-R}

We start with a simple lemma about blow-up of solutions to an ODE with
a quadratic nonlinearity (see, e.g., \cite{Alinhac:blowup}, p.~72 or
\cite{Hormander:nonlin-hyp}, Lemma~1.3.2).
\begin{lemma}
  \label{la:balance-ode}
  Let $v(t)$ be a solution on $[t_0,t_1)$ of
  \begin{equation}\label{eq:balance-ode-dvdt}
    \frac{\rmd{v}}{\rmd{t}}
    = A_2(t) v^2 + A_1(t) v + A_0(t),
    \qquad
    v(t_0) = v_0,
  \end{equation}
  where $A_2$, $A_1$ and $A_0$ are continuous and bounded on
  $[t_0,t_1)$ with $A_2\ge0$. Put
  \begin{equation}
    K_1(t) = \int_{t_0}^t A_1 \,\rmd{t},
    \qquad
    K_0(t) = \int_{t_0}^t \abs{A_0} \rme^{-K_1} \,\rmd{t},
    \qquad
    K_2(t) = \int_{t_0}^t A_2 \rme^{K_1} \,\rmd{t}.
  \end{equation}
  If $v_0>K_0(t_1)$ then
  \begin{equation}\label{eq:balance-ode-v0cond}
    K_2(t_1) < \bigl( v_0 - K_0(t_1) \bigr)^{-1}
  \end{equation}
  and we have the following lower bound on $v$ in $[t_0,t_1)$:
  \begin{equation}
    \label{eq:balance-ode-vbound}
    \rme^{-K_1(t)} v(t)
    \ge K_0(t_1) - K_0(t)
        + \Bigl[ \bigl(v_0-K_0(t_1)\bigr)^{-1} - K_2(t) \Bigr]^{-1}.
  \end{equation}
\end{lemma}
\begin{proof}
  The linear term can be dealt with by putting $V=v\rme^{-K_1}$,
  which transforms \eqref{eq:balance-ode-dvdt} into
  \begin{equation}\label{eq:balance-ode-dVdt}
    \frac{\rmd{V}}{\rmd{t}}
    = A_2\rme^{K_1} V^2 + A_0\rme^{-K_1},
    \qquad
    V(t_0) = v_0,
  \end{equation}
  Let $W$ be the solution of
  \begin{equation}
    \frac{\rmd{W}}{\rmd{t}}
    = A_2\rme^{K_1} \bigl( W - K_0(t_1) + K_0(t) \bigr)^2
    - \abs{A_0}\rme^{-K_1},
    \qquad
    W(t_0) = v_0,
  \end{equation}
  which can be found explicitly as
  \begin{equation}
    W(t) = K_0(t_1) - K_0(t)
         + \Bigl[ \bigl( v_0 - K_0(t_1) \bigr)^{-1} - K_2(t) \Bigr]^{-1}.
  \end{equation}
  We want to show \eqref{eq:balance-ode-vbound}, i.e., $V\ge{W}$.
  Now
  \begin{equation}\label{eq:balance-ode-dwdt}
    \frac{\rmd{W}}{\rmd{t}} \le A_2\rme^{K_1} W^2 + A_0\rme^{-K_1},
  \end{equation}
  so subtracting~\eqref{eq:balance-ode-dVdt}
  from~\eqref{eq:balance-ode-dwdt} gives
  \begin{equation}
    \frac{\rmd}{\rmd{t}} (W - V) \le A_2\rme^{K_1} (W + V)(W - V),
  \end{equation}
  and a Gronwall estimate implies that $W-V\le0$ as long as $W+V\ge0$.
  Whenever $W-V\le0$, $W+V\ge2W$, so $V\ge{W}$ as long as $W\ge0$. By
  definition, $W\ge0$ when $K_2(t)<(v_0-K_0(t_1))^{-1}$. But
  $W\to\infty$ as $K_2(t)\to(v_0-K_0(t_1))^{-1}$, so since $V$ is bounded on
  $[t_0,t_1)$ we must have $K_2(t_1)<(v_0-K_0(t_1))^{-1}$.
\end{proof}

Note that in \cite{Alinhac:blowup} and \cite{Hormander:nonlin-hyp},
the factors involving $K_1$ are moved outside the integrals. We
avoid this since we want to keep the simple form of
\eqref{eq:balance-ode-vbound}.

\begin{lemma}
  \label{la:balance-R}
  Let $\U$ be an open set in $\reals^2$ and put
  $\Omega=[t_0,t_1)\times\sphere\times\U$.  Assume that
  \begin{enumerate}
  \item\label{ass:balance-R-rsbound}%
    $(r,s)$ remains in $\U$ for $t\in[t_0,t_1)$,
  \item\label{ass:balance-R-ebound}%
    $e_0\^0 + \kappa e_1\^0$ is continuous and positive on $\Omega$,
  \item\label{ass:balance-R-Abound}%
    $A_2$, $A_1$ and $A_0$, given by \eqref{eq:balance-Adef}, are
    continuous with $A_2>0$ and
    \begin{equation}
      \label{eq:balance-R-Cdef}%
      C_2 = \inf_\U A_2,
      \qquad
      C_1 = \sup_{\Omega} \abs{A_1}
      \qquad\text{and}\qquad
      C_0 = \sup_{\Omega} \abs{A_0}
    \end{equation}
    are all finite with $C_2>0$, and
  \item\label{ass:balance-R-phi}%
    $\phi$ is bounded on $\Omega$.
  \end{enumerate}
  If
  \begin{equation}
    \label{eq:balance-R-R0bound}
    R(t_0,x) \ge \phi(t_0,x,r_0(x),s_0(x)) + C_0 \chi(t_1) 
               + \bigl( C_2 \chi(t_1) \bigr)^{-1} + C_1/C_2
  \end{equation}
  for some $x\in\sphere$, where
  \begin{equation}
    \label{eq:balance-R-chidef}%
    \chi(t_1) =
    \begin{cases}
      (\rme^{C_1\xi(t_1)}-1)/C_1 & \text{if $C_1>0$}, \\
      \xi(t_1)                   & \text{if $C_1=0$}
    \end{cases}
  \end{equation}
  and $\xi(t_1)$ is the parameter value along the characteristic
  $\varrho_x$ corresponding to the time $t_1$, then $R$ cannot be
  bounded on $[t_0,t_1]\times\sphere$.
\end{lemma}

\begin{proof}
  Assumption~\ref{ass:balance-R-rsbound} ensures that the other
  assumptions apply along the characteristic $\varrho_x$ for
  $t\in[t_0,t_1]$.  Along $\varrho_x$, the parameter $\xi$ satisfies
  $\rmd{t}/\rmd\xi=e_0\^0+\kappa e_1\^0$, so it follows from
  assumption~\ref{ass:balance-R-ebound} that $\xi$ is a continuously
  differentiable and increasing function of $t$ along $\varrho_x$.
  
  Because of assumption~\ref{ass:balance-R-Abound} and assuming that
  $R-\phi$ is bounded, we can apply Lemma~\ref{la:balance-ode} to
  \eqref{eq:balance-R-phi} along $\varrho_x$, giving
  \begin{equation}
    R(t_0,x)-\phi(t_0,x,r_0(x),s_0(x))
    < K_0\bigl(\xi(t_1)\bigr) + K_2\bigl(\xi(t_1)\bigr)^{-1}.
  \end{equation}
  If $C_1>0$, we have
  \begin{equation}
    \abs{K_1} \le C_1 \xi(t),
    \qquad
    K_0 \le \frac{C_0}{C_1} (\rme^{C_1 \xi(t_1)} - 1)
    \qquad\text{and}\qquad
    K_2 \ge \frac{C_2}{C_1} (1 - \rme^{-C_1 \xi(t_1)}),
  \end{equation}
  so
  \begin{equation}
    R(t_0,x)-\phi(t_0,x,r_0(x),s_0(x))
    < C_0 \chi(t_1) + \bigl( C_2 \chi(t_1) \bigr)^{-1} + C_1/C_2,
  \end{equation}
  where $\chi(t_1) = (\rme^{C_1\xi(t_1)}-1)/C_1$. When $C_1=0$ the
  same inequality holds but with $\chi(t_1)=\xi(t_1)$. This
  contradicts the assumptions of the lemma and so in fact $R-\phi$ is
  unbounded. It follows that $R$ must blow up since $\phi$ does not.
\end{proof}

\begin{theorem}
  \label{th:balance-blowup-rs}
  Let $\U_\delta$ and $\Omega_\delta$ be as
  in~\eqref{eq:balance-UOmegadef}.  Assume that
  \begin{enumerate}
    \renewcommand{\theenumi}{\normalfont\Roman{enumi}}
  \item\label{ass:balance-rs-kappalambda}%
    $\lambda$, $\lambda_r$, $\lambda_s$, $\kappa_r$ and $\kappa_{s}$
    are continuous and bounded on $\U_\delta$, with $\kappa_r$ and
    $\kappa-\lambda$ bounded away from $0$,
  \item\label{ass:balance-rs-fg}%
    $f$, $f_r$, $f_s$, $f_{se_0}$, $f_{e_1}$ and $g$ are continuous and
    bounded on $\Omega_\delta$,
  \item\label{ass:balance-rs-e}%
    $e_0\^0+{\kappa}e_1\^0$ and $e_0\^0+{\lambda}e_1\^0$ are
    continuous, bounded, positive and bounded away from $0$ on
    $\Omega_\delta$, and
  \item\label{ass:balance-rs-c}%
    $c^0\_{01}$ and $c^1\_{01}$ are continuous and bounded on
    $[t_0,t_1)\times\sphere$.
  \end{enumerate}
  Suppose also that
  \begin{equation}
    \label{eq:balance-rs-e1r}
    e_1(r)(t_0,x)
    \ge \rme^{-h(r_0,s_0)}
         \Bigl(
           \phi(t_0,x,r_0,s_0)
           + C_0 \chi(\hat{t}_1)
           + \bigl( C_2 \chi(\hat{t}_1) \bigr)^{-1}
           + C_1/C_2
         \Bigr)
  \end{equation}
  for some $x\in\sphere$, where $\hat{t}_1 = \min\{t_0 +
  2^{-1/2}\delta EF^{-1}, t_1\}$, $F$ and $E$ are given by
  \eqref{eq:balance-FEdef}, $h$ is given by~\eqref{eq:balance-hdef},
  $\phi$ by~\eqref{eq:balance-phidef}, $C_i$
  by~\eqref{eq:balance-R-Cdef} and $\chi$
  by~\eqref{eq:balance-R-chidef}.
  Then there is no smooth solution of
  \eqref{eq:balance-laws} on $[t_0,t_1]$ with initial data
  $(r_0,s_0)$.

  Moreover, the right hand side of~\eqref{eq:balance-rs-e1r} can be
  estimated in terms of $\norm{r_0}$, $\norm{s_0}$, $t_0$, $t_1$, $x$,
  $\delta$ and the bounds in
  assumptions~\ref{ass:balance-rs-kappalambda}--\ref{ass:balance-rs-c}.
\end{theorem}

\begin{proof}
  First of all, that $(r,s)\in\U_\delta$ for $t\in[t_0,\hat{t}_1)$
  follows from Lemma~\ref{la:balance-rsbound}, which applies because
  of assumptions~\ref{ass:balance-rs-fg} and \ref{ass:balance-rs-e}.
  Thus condition~\ref{ass:balance-R-rsbound} of
  Lemma~\ref{la:balance-R} holds. Condition~\ref{ass:balance-R-ebound}
  follows directly from assumption~\ref{ass:balance-rs-e}.

  To simplify the presentation we will denote the supremum norms over
  $\sphere$, $\U_\delta$ and $\Omega_\delta$ with the same symbol
  $\norm{\cdot}$.  From the definition~\eqref{eq:balance-hdef} and
  Lemma~\ref{la:balance-gamma}, $h$ is continuous on $\U_\delta$ with
  \begin{equation}
    \label{eq:balance-rs-h}
    \norm{h}
    \le \norm{(\kappa-\lambda)^{-1}}
        \norm{\kappa_s}
        (2\norm{s_0} + \delta).
  \end{equation}
  It follows that $h$ is bounded because of
  assumption~\ref{ass:balance-rs-kappalambda}.
  Differentiating~\eqref{eq:balance-hdef} by $r$ and performing a
  partial integration to get rid of the mixed second partial
  derivative $\kappa_{rs}$ gives an estimate
  \begin{equation}
  \label{eq:balance-rs-h_r}
  \begin{split}
    \norm{h_r}
    \le\ &\norm{(\kappa-\lambda)^{-1}}
          \bigl(
            2\norm{\kappa_r}
           + \norm{\kappa_s}
          \bigr) \\
    +\   &\norm{(\kappa-\lambda)^{-1}}^2
           \bigl(
             \norm{\kappa_r}\norm{\lambda_s}
           + \norm{\kappa_s}\norm{\lambda_r}
           \bigr)
        (2\norm{s_0}+\delta),
  \end{split}
  \end{equation}
  so it follows from assumption~\ref{ass:balance-rs-kappalambda} that
  $h_r$ is bounded as well.
  
  From the definition~\eqref{eq:balance-phidef} and
  assumptions~\ref{ass:balance-rs-kappalambda} and
  \ref{ass:balance-rs-fg}, $\phi$ is continuous on $\Omega_\delta$ with
  \begin{equation}
    \label{eq:balance-rs-phi}
    \norm{\phi}
    \le \norm{(\kappa-\lambda)^{-1}}
        \rme^{\norm{h}}
        \norm{f_s}
        (2\norm{s_0} + \delta).
  \end{equation}
  Differentiating $\phi$ with respect to $r$ and performing a partial
  integration to avoid the term involving $f_{rs}$ gives an estimate
  \begin{equation}
  \label{eq:balance-rs-phir}
  \begin{split}
    \norm{\phi_r}
    \le \rme^{\norm{h}} \norm{(\kappa-\lambda)^{-1}}
        \Bigl[
          2 \norm{f_r} + \norm{f_s}
        + \bigl(
            \norm{f_r}\norm{H_s}
          + \norm{f_s}\norm{H_r}
          \bigr)
          (2\norm{s_0}+\delta)
        \Bigr],
  \end{split}
  \end{equation}
  where $H=h-\ln(\kappa-\lambda)$. Differentiating $\phi$ with respect
  to $e_0$ gives the estimate
  \begin{equation}
    \label{eq:balance-rs-phie0}
    \norm{\phi_{e_0}}
    \le \rme^{\norm{h}} \norm{(\kappa-\lambda)^{-1}}
        \norm{f_{se_0}} (2\norm{s_0}+\delta),
  \end{equation}
  while for $\phi_{e_1}$ we perform a partial integration before
  estimating, giving
  \begin{equation}
    \label{eq:balance-rs-phie1}
    \norm{\phi_{e_1}}
    \le \rme^{\norm{h}} \norm{(\kappa-\lambda)^{-1}}
        \norm{f_{e_1}}
        \bigl(
          2 + \norm{H_s} (2\norm{s_0}+\delta)
        \bigr).
  \end{equation}
  By assumptions~\ref{ass:balance-rs-kappalambda} and
  \ref{ass:balance-rs-fg}, $\phi$, $\phi_r$, $\phi_{e_0}$ and
  $\phi_{e_1}$ are bounded. We conclude that all terms
  in~\eqref{eq:balance-Adef} are continuous and bounded.
  
  Now $\inf_{\U_\delta}{A_2}>0$ because of
  assumption~\ref{ass:balance-rs-kappalambda}, so all conditions of
  Lemma~\ref{la:balance-R} are met. Since $R=\rme^{h}e_1(r)$, $e_1(r)$
  cannot be bounded on the characteristic $\varrho_x$ if
  \eqref{eq:balance-rs-e1r} holds.

  Finally, the only quantity in the right hand side
  of~\eqref{eq:balance-rs-e1r} whose dependence on the estimates has
  not been established by the arguments above is $\chi(\hat{t}_1)$.
  From the definition~\eqref{eq:balance-FEdef} of $E$ and $F$,
  $\hat{t}_1$ can be estimated both from below and above by $t_0$,
  $t_1$, $x$, $\delta$, $\norm{r_0}$, $\norm{s_0}$ and the bounds in
  assumptions~\ref{ass:balance-rs-kappalambda}--\ref{ass:balance-rs-c}.
  That the same holds for $\chi(\hat{t}_1)$ follows from the
  definition~\eqref{eq:balance-R-chidef} of $\chi$ together with the
  estimate of $A_1$ above.
\end{proof}

The condition on $f_{se_0}$ can of course be replaced by a condition on
$f_{e_0}$ by performing a partial integration as we did with $h_r$,
$f_{rs}$ and $f_{se_1}$. The reason for not doing so will become
evident below. Note also that there is no explicit dependence of $f_{e_0}$
in~\eqref{eq:balance-Adef}.

Let us define the blowup time as 
\begin{equation}
  t_* = \sup\{ t>t_0;\; (r,s) \text{ is smooth on } [t_0,t]\}.
\end{equation}
Then we also have the following.
\begin{corollary}
  Under the conditions of Theorem~\ref{th:balance-blowup-rs}, the
  blowup time $t_*$ satisfies $\chi(t_*)\le\chi_*$ where $\chi_*$ is
  the smaller root of
  \begin{equation}
    e_1(r)(t_0,x)
    = \rme^{-h(r_0,s_0)}
        \Bigl(
          \phi(t_0,x,r_0,s_0)
          + C_0 \chi
          + \bigl( C_2 \chi \bigr)^{-1}
          + C_1/C_2
        \Bigr).
  \end{equation}
\end{corollary}


\section{Blowup in spacetime}
\label{sec:blowup}

We will now combine the results of the preceding sections to show
finite life span of solutions to the Einstein-Euler equations. We
first need to control higher derivatives of some geometric quantities,
given bounds on the matter variables.

\begin{lemma}
  \label{la:blowup-geobounds}
  If the conditions of Lemma~\ref{la:Euler-smooth} are satisfied and
  $r$ and $s$ are bounded with $r+s>2\varphi_-$, the following
  quantities are bounded:
  \begin{equation}
    \label{eq:blowup-geobounds}
    K', A'', \alpha'', \d_t{K}, \beta'' \text{ and } \d_t{A'}.
  \end{equation}
  The bounds involve only those in~\eqref{eq:Einstein-bounds} and the
  a priori bounds on $r$ and $s$.
\end{lemma}

\begin{proof}
  From Lemma~\ref{la:Euler-smooth}, $\rho$, $j$ and $S$ can be bounded
  in terms of $r$ and $s$ for a given equation of state. Using these
  bounds together with~\eqref{eq:Einstein-bounded-vars} in
  \eqref{eq:Einstein-momentum}, \eqref{eq:Einstein-Hamiltonian},
  \eqref{eq:Einstein-lapse} and \eqref{eq:Einstein-dKdt} immediately
  gives bounds on $K'$, $A''$, $\alpha''$ and $\d_t{K}$.
  Differentiating~\eqref{eq:Einstein-dadt} by $x$ gives $\beta''$
  expressed in $\alpha$, $\alpha'$, $K$, $K'$ and $t$, so it is
  bounded too. Finally, differentiating \eqref{eq:Einstein-dAdt} by
  $x$ gives a bound on $\d_t{A'}$ in terms of the previously obtained
  bounds.
\end{proof}

The previous results can now be combined to show a blow up result for
the Einstein-Euler equations in plane symmetry.

\begin{theorem}
  \label{th:blowup-theorem}
  Let $t_0<t_1<0$ and suppose that the conditions of
  Lemma~\ref{la:Euler-smooth} hold.  Then there are smooth initial
  data on the surface $t=t_0$ such that the corresponding smooth
  solution of the Einstein-Euler equations~\eqref{eq:Einstein} and
  \eqref{eq:Euler1} does not extend beyond $t=t_1$.
\end{theorem}

\begin{proof}
  Theorem~\ref{th:Einstein-bounds} shows that all the geometric
  quantities in~\eqref{eq:Einstein-bounded-vars} are bounded in terms of the
  quantities in~\eqref{eq:Einstein-bounds}. Let $\U_\delta$ and
  $\Omega_\delta$ be as in~\eqref{eq:balance-UOmegadef}, with $\delta$
  as in~\eqref{eq:balance-deltabound}. Then $\U_\delta$ and
  $\Omega_\delta$ depend on $r_0$, $s_0$, $t_0$, $t_1$ and $\delta$,
  and $\delta$ is limited by $\varphi_0-\varphi_-$, which in turn
  depends only on the equation of state and the minimum of $r_0+s_0$.
  By construction, $r$ and $s$ can be bounded on $\Omega_\delta$ in
  terms of $\delta$ and the maximum values of $r_0$ and $s_0$. It then
  follows from Lemma~\ref{la:blowup-geobounds} that the geometric
  quantities~\eqref{eq:blowup-geobounds} also are bounded, in
  terms of~\eqref{eq:Einstein-bounds} and
  \begin{equation}
    \label{eq:blowup-rsbounds}
    \delta, \norm{r_0}, \norm{s_0}.
  \end{equation}
  
  We need to establish the bounds in
  Theorem~\ref{th:balance-blowup-rs}. First
  Theorem~\ref{th:Einstein-bounds} and Lemma~\ref{la:blowup-geobounds}
  give that
  \begin{equation}
    \label{eq:blowup-e00-e10}
    e_0\^0 + {\kappa} e_1\^0
    = e_0\^0 + {\lambda} e_1\^0
    = \alpha^{-1}
  \end{equation}
  are continuous, positive and bounded away from $0$ and
  \begin{equation}
    \label{eq:blowup-c001-c101}
    c^0\_{01} = \alpha^{-1} e_1(\alpha)
    \quad\text{and}\quad
    c^1\_{01} = \alpha^{-1} A e_1(\beta) - A^{-1} e_0(A)  
  \end{equation}
  are continuous and bounded in terms of~\eqref{eq:Einstein-bounds}.
  
  From the definition of $u$ and the construction of $\U_\delta$,
  $\ln(1-u^2)$, $(1-u^2)^{-1}$ and $(1\pm{uw})^{-1}$ can be bounded in
  terms of~\eqref{eq:blowup-rsbounds}. Also,
  \begin{equation}
    \frac12\inf_{\sphere}\{r_0+s_0\} -\delta
    < \varphi
    < \frac12\sup_{\sphere}\{r_0+s_0\} + \delta,
  \end{equation}
  so for a given equation of state, Lemma~\ref{la:Euler-smooth} gives
  that $w^{-1}$, $(1-w^2)^{-1}$ and $\frac{\rmd{w}}{\rmd\varphi}$ are
  bounded in terms of~\eqref{eq:blowup-rsbounds} and
  \begin{equation}
    \label{eq:blowup-infrsbound}
    (\varphi_0 - \varphi_- - \delta)^{-1}
    = \Bigl( \frac12\inf_{\sphere}\{r_0+s_0\} - \varphi_- - \delta \Bigr)^{-1}.
  \end{equation}

  The balance law coefficients can be read off directly
  from~\eqref{eq:Euler-balance}. They are
  \begin{subequations}
    \label{eq:blowup-balancecoeff}
    \begin{align}
      \label{eq:blowup-kappa}
      \kappa(r,s) &= \frac{u+w}{1+uw}, \\
      \label{eq:blowup-lambda}
      \lambda(r,s) &= \frac{u-w}{1-uw}, \\
      \label{eq:blowup-f}
      f(t,x,r,s) &= \frac{t w + K u - 2 u w A^{-1} e_1(A)}{1+uw}
                  - \alpha^{-1} e_1(\alpha), \\
      \label{eq:blowup-g}
      g(t,x,r,s) &= \frac{t w - K u - 2 u w A^{-1} e_1(A)}{1-uw}
                  + \alpha^{-1} e_1(\alpha).
    \end{align}
  \end{subequations}
  From the previously obtained bounds it follows that $\kappa$ and
  $\lambda$ are continuous on $\U_\delta$ and can be bounded in terms
  of~\eqref{eq:blowup-rsbounds}. Differentiating and using that
  $w_r=w_s=\frac12\frac{\rmd{w}}{\rmd\varphi}$ and
  $u_r=-u_s=\frac12(1-u^2)$ we get
  \begin{equation}
    \kappa_r 
    = \frac{1 - u^2}{2 (1 + uw)^2}
      \Bigl(1 - w^2 + \frac{\rmd{w}}{\rmd\varphi} \Bigr),
  \end{equation}
  with similar expressions for $\kappa_s$, $\lambda_r$ and
  $\lambda_s$.  These quantities are continuous on $\U_\delta$ and can
  be bounded in terms of~\eqref{eq:blowup-rsbounds} and
  \eqref{eq:blowup-infrsbound}. Moreover, $\kappa_r$ and
  \begin{equation}
    \kappa - \lambda = \frac{2 w (1 - u^2)}{1 - u^2 w^2}
  \end{equation}
  are positive, and $\kappa_r^{-1}$ and $(\kappa-\lambda)^{-1}$ are
  bounded in terms of~\eqref{eq:blowup-rsbounds} and
  \eqref{eq:blowup-infrsbound}.

  From~\eqref{eq:blowup-balancecoeff}, $f$ and $g$ are continuous on
  $\Omega_\delta$ and bounded in terms of~\eqref{eq:Einstein-bounds}
  and \eqref{eq:blowup-rsbounds}. Since
  $w_r=w_s=\frac12\frac{\rmd{w}}{\rmd\varphi}$ and
  $u_r=-u_s=\frac12(1-u^2)$, the same holds for their $r$ and $s$
  derivatives.

  Finally the quantities appearing in $f_{se_0}$ and $f_{e_1}$ not
  present in $f$ are $\d_t{K}$, $K'$, $\d_t{A'}$, $A''$ and
  $\alpha''$. It follows from Lemma~\ref{la:blowup-geobounds} that
  $f_{se_0}$ and $f_{e_1}$ are continuous and bounded in terms
  of~\eqref{eq:Einstein-bounds}, \eqref{eq:blowup-rsbounds} and
  \eqref{eq:blowup-infrsbound}. Note that $f_{se_0}$ does not depend
  on $\d_t\alpha'$ since the term in~\eqref{eq:blowup-f} containing
  $\alpha$ is independent of $s$.  This is the reason for not
  replacing $f_{se_0}$ by $f_{e_0}$ in
  Theorem~\ref{th:balance-blowup-rs}.

  Now all the conditions of Theorem~\ref{th:balance-blowup-rs} have
  been established and we can conclude that if the initial data
  satisfies~\eqref{eq:balance-rs-e1r} for some $x$, there is no smooth
  solution on $[t_0,t_1]$.
  By definition, $e_1(r)=A^{-1}\d_x{r}$. Since all quantities
  in~\eqref{eq:balance-rs-e1r} except $\d_x{r}$ can be estimated in
  terms of the geometric initial data~\eqref{eq:Einstein-bounds} and
  the matter initial data~\eqref{eq:blowup-rsbounds} and
  \eqref{eq:blowup-infrsbound}, choosing initial data with
  sufficiently large gradient $\d_x{r}$ will prevent a
  smooth extension up to $t=t_1$.

\end{proof}

The question, how big the class of solutions of the constraints is which 
satisfy the hypotheses of the theorem will not be treated in great
generality here. It will, however, be shown that there are some data 
which do so for which the time of existence is arbitrarily short. Consider
data for which $u$ is identically zero. Then a particular solution of equation 
(\ref{eq:Einstein-momentum}) is given by $K=t/3$. Substituting this into the 
remaining constraint (\ref{eq:Einstein-Hamiltonian}) gives 
$(\sqrt{A})''=-\frac18 A^{5/2}(-\frac23+16\pi\mu)$. The equation
to be solved is a special case of one treated in 
\cite{Isenberg-Rendall:cosmo-not-CMC}
where the existence of a solution $A$ was shown by the method of sub- and 
supersolutions. This method also gives uniform bounds for the solution
which in particular shows that the derivative $\mu'$ at a point can be made 
arbitrarily large within a family of solutions while maintaining a fixed 
$L^\infty$ bound for $A$. At the same time fixed $L^\infty$ bounds for $\mu$ 
and $\mu^{-1}$ can be maintained. This suffices to show that $e_1(r)$
becomes arbitrarily large within the family and provides the desired
data for which the time of existence is arbitrarily short.


\section{An extension result}
\label{sec:extension}

In the previous section we established that if the initial data has
sufficiently large gradients, there is no smooth solution of the
Einstein-Euler equations beyond a certain time. It seems quite
plausible that the obstruction to extending the solution is the blowup
of first derivatives of the matter variables, although it must be
pointed out that we have not shown that. To investigate this further,
we will show that if the first derivatives of the matter variables are
bounded then the spacetime can be extended. The argument will be
similar to that in~\cite{Rendall:crushing-sings}. First we need bounds
on $r$ and $s$, which can be established under a mild extra condition
on the equation of state.

\begin{theorem}
  \label{th:extension-rsbound}
  Assume that in addition to the conditions in
  Lemma~\ref{la:Euler-smooth}, $\rmd{p}/\rmd\mu$ is bounded away from
  $1$. Then the following quantities are bounded:
  \begin{equation}
    \label{eq:extension-rsbound}
    r, s, \rho, j, S.
  \end{equation}
\end{theorem}

\begin{proof}
  The extra condition and Lemma~\ref{la:Euler-smooth} implies that the
  factors in the denominators of~\eqref{eq:blowup-balancecoeff} are
  positive and bounded away from $0$. Thus the right hand sides
  of~\eqref{eq:balance-laws} can be bounded in terms of the geometric
  quantities~\eqref{eq:Einstein-bounded-vars}.
  Integrating~\eqref{eq:balance-laws} shows that $r$ and $s$ are
  bounded along the characteristics. We may pass from these estimates
  to estimates in time because of~\eqref{eq:blowup-e00-e10}. Finally,
  $\rho$, $j$ and $S$ must also be bounded because of
  Lemma~\ref{la:Euler-smooth}.
\end{proof}

\begin{lemma}
  \label{la:extension-dx++geom}
  Under the conditions of Lemma~\ref{la:Euler-smooth}, if all spatial
  derivatives of order up to $n$ of the geometric
  quantities~\eqref{eq:Einstein-bounded-vars} and the matter
  quantities~\eqref{eq:extension-rsbound} are bounded, the spatial
  derivatives of order up to $n+1$ of the geometric
  quantities~\eqref{eq:Einstein-bounded-vars} are bounded.
\end{lemma}

\begin{proof}
  Note first that some of the spatial derivatives are already included
  in~\eqref{eq:Einstein-bounded-vars}. As was shown in the proof of
  Lemma~\ref{la:blowup-geobounds}, we can solve~\eqref{eq:Einstein}
  for the remaining quantities $\alpha''$, $A''$, $\d_t{A'}$, $K'$ and
  $\beta''$. Differentiating the resulting equations $n$ times gives
  the desired bounds.
\end{proof}

\begin{lemma}
  \label{la:extension-dx++matter}
  Suppose that the conditions of
  Lemma~\ref{la:extension-dx++geom} hold, $\rmd{p}/\rmd\mu$
  is bounded away from $1$ and $\d_x{r}$ and $\d_x{s}$ are bounded.
  Then the spatial derivatives of the matter
  quantities~\eqref{eq:extension-rsbound} of order up to $n+1$ are
  bounded.
\end{lemma}

\begin{proof}
  Note first that because of the definition~\eqref{eq:Einstein-e} of
  $e_1$, using $e_1^n$ instead of $\d_x^n$ introduces spatial
  derivatives of $A^{-1}$ of order at most $n-1$, which are bounded by
  assumption.
  
  Put $r_n=e_1^n(r)$ and $s_n=e_1^n(s)$.
  Applying~\eqref{eq:balance-Dpluse1} to $r_{n+1}$ gives
  \begin{equation}
    \label{eq:extension-Dplusrn}
    \Dplus r_{n+1}
    = (e_1 - c^0\_{10}) \Dplus r_n
      - \bigl(
          \kappa_r e_1(r) - \kappa_s e_1(s)
          + c^1\_{10} - \kappa c^0\_{10}
        \bigr) r_{n+1}.
  \end{equation}
  If we can show that~\eqref{eq:extension-Dplusrn} is a linear
  equation in $\Dplus{r_{n+1}}$ with the coefficients given by smooth
  functions of spatial derivatives of $r$ and $s$ of order up to $n$
  and of the geometric quantities~\eqref{eq:Einstein-bounded-vars} of
  order up to $n+1$, we would be done because of
  Lemma~\ref{la:extension-dx++geom}.
  
  Putting $n=1$ and using the expression~\eqref{eq:balance-Dpluse1r}
  for $\Dplus r_1$ shows that $\Dplus r_2$ is linear in $r_2$, and the
  coefficients have the right dependence because
  of~\eqref{eq:blowup-balancecoeff} and \eqref{eq:blowup-c001-c101}.
  For $n>2$ the coefficients of $\Dplus{r_n}$ are differentiated by
  $x$ at most once in~\eqref{eq:extension-Dplusrn} so the result
  follows by induction. The case with $s$ is completely analogous.
\end{proof}

\begin{lemma}
  \label{la:extension-dt++matter}
  Under the conditions of Lemma~\ref{la:extension-dx++matter}, if all
  derivatives of the geometric quantities~\eqref{eq:Einstein-bounded-vars}
  and the matter quantities~\eqref{eq:extension-rsbound} of the form
  $\d_t^k\d_x^n$ with $n\ge0$ and $0\le{k}\le{m}$ are bounded then the
  $\d_t^{m+1}\d_x^n$ derivatives of the matter
  quantities~\eqref{eq:extension-rsbound} are bounded.
\end{lemma}

\begin{proof}
  As was the case with $e_1$ and $\d_x$, we may use $e_0$ instead of
  $\d_t$ because of~\eqref{eq:Einstein-e}. The balance
  laws~\eqref{eq:balance-laws} can be used to bound an extra time
  derivative by introducing an extra space derivative, which is
  bounded by Lemma~\ref{la:extension-dx++matter}.
\end{proof}

\begin{lemma}
  \label{la:extension-dt++geom}
  Under the conditions of Lemma~\ref{la:extension-dx++matter}, if
  \begin{enumerate}
  \item all derivatives of the geometric
    quantities~\eqref{eq:Einstein-bounded-vars} and the matter
    quantities~\eqref{eq:extension-rsbound} of the form $\d_t^k\d_x^n$
    with $n\ge0$ and $0\le{k}\le{m}$ are bounded, and
  \item all derivatives of the matter
    quantities~\eqref{eq:extension-rsbound} of the form
    $\d_t^{m+1}\d_x^n$ are bounded,
  \end{enumerate}
  then the derivatives of the geometric
  quantities~\eqref{eq:Einstein-bounded-vars} of the form $\d_t^{m+1}\d_x^n$
  are bounded.
\end{lemma}

\begin{proof}
  See~\cite{Rendall:crushing-sings}.
\end{proof}

Putting together Theorem~\ref{th:extension-rsbound} and
Lemma~\ref{la:extension-dx++geom}--~\ref{la:extension-dt++geom} we see
that under the hypotheses of Lemma~\ref{la:extension-dx++matter}, all
derivatives of the geometric and matter quantities are bounded. When
all the derivatives are bounded on a given interval, the solution can
be smoothly extended to the closure of that interval. This results in
a new initial data set and applying the local existence and uniqueness
result from~\cite{Rendall:crushing-sings} gives an extension of the
solution. We thus have the following theorem.

\begin{theorem}
  \label{th:extension-exists}
  Let a solution of the Einstein equations with plane symmetry be
  given. Suppose that when coordinates are chosen which cast the
  metric into the form~\eqref{eq:Einstein-ds2} with constant mean
  curvature time slices, the time coordinate takes all values in the
  finite interval $[t_0,t_1)$ with $t_1<0$.  Assume that the equation
  of state satisfies the conditions in Lemma~\ref{la:Euler-smooth},
  and that $\rmd{p}/\rmd\mu$ is bounded away from $1$. If the first
  spatial derivatives of the energy density $\mu$ and the velocity
  parameter $u$ are bounded on $[t_0,t_1)$, the solution can be
  extended beyond $t=t_1$.
\end{theorem}


\section{Discussion}
\label{sec:discussion}

In this paper it has been shown that plane symmetric classical
solutions of the Einstein-Euler system can break down in finite time
under some general assumptions on the equation of state. Time here is
measured with respect to a CMC foliation. That the formation of
singularities is due to the fluid and not just to a breakdown of the
CMC foliation is shown by the following fact. If instead of the Euler
equations we describe the matter content of spacetime by a
collisionless gas satisfying the Vlasov equations and maintain all
other assumptions the solutions exist globally in the
future~\cite{Rendall:crushing-sings}. In the case of the Euler
equations for a fluid without pressure (dust) a non-existence result
was proved in~\cite{Rendall:exist-nonexist-CMC}.  The proofs of all
these results have common elements. It is shown that the spacetime
geometry has certain good properties right up to the singularity.
Using this it is shown that when there are no singularities in flat
spacetime there are also none after coupling to the Einstein equations
(Vlasov) while in the cases there are singularities (Euler with
pressure, dust) the mechanisms of blow-up in the flat space case
(formation of shocks, blow-up of the density) provide a reliable guide
as to what quantities are bounded or not and how to prove it in the
case of coupling to the Einstein equations. 

A family of data along which the time of existence becomes arbitrarily small
can be used to show that the results of \cite{Isenberg-Rendall:cosmo-not-CMC} 
on solutions
of the Einstein-dust system generalize to the case of fluids with non-zero 
pressure. The conclusion is that there are data for the Einstein-Euler
equations with an equation of state of the kind considered in the present
paper on a CMC hypersurface such that the corresponding maximal Cauchy
development cannot be foliated by compact CMC hypersurfaces. To see that
the proofs of \cite{Isenberg-Rendall:cosmo-not-CMC} can be adapted note first 
that under 
the assumption $u=0$ the constraints for the fluid with pressure are 
identical to those for dust so that many of the arguments can be taken over 
directly. Cauchy stability for the Einstein-Euler system with 
$0<\frac{dp}{d\mu}<1$, 
which is necessary for the argument, follows from the fact that it can be 
written in symmetric hyperbolic form so that standard results apply. See
for instance \cite{Rendall:book}, in particular p. 140. The final element 
which is required is to know that the argument for breakdown of the 
solution can be localized in space. Since the contradiction to existence
on a certain time interval is obtained by an analysis of the evolution of
certain quantities along a single characteristic it is only necessary to
show that in a short time interval the characteristic can only move a
short (coordinate) distance in space. This follows immediately from the 
available a priori bounds for the geometry.

The assumptions on the equation of state in
Theorem~\ref{th:blowup-theorem} are compatible with cases of physical
relevance. Consider, for instance, the relativistic polytrope with
\begin{equation}
  \mu=m+Knm^{1+\frac1n},\ \ \ p=Km^{1+\frac1n}
\end{equation}
where $n>1$ and $K>0$ are constants and $m$ the rest mass density. In
this case all the assumptions are satisfied. They are also satisfied
for an equation of state leading to a linear relation $p=K\mu$ with
$K>0$, as considered in~\cite{Smoller-Temple:relativistic-Euler}
and~\cite{Barnes-etal:Glimm-Gowdy}. This is true in spite of the fact
that if a full thermodynamic treatment of this case is attempted the 
internal energy is negative at low densities and so its physical 
interpretation is questionable.

In the case of Theorem~\ref{th:blowup-theorem} the density remains
bounded on the maximal interval of existence of a classical solution.
This cannot be expected to hold for dust, although it is not actually
proved in~\cite{Rendall:exist-nonexist-CMC} that the density is
unbounded.  Thus some restrictions on the equation of state are to be
expected. It is probable that blow-up of the density occurs when the
equation of state is such that the pressure is a bounded function of
the density (cf.~\cite{Yodzis-Seifert-zumHagen:naked-II}). In
Theorem~\ref{th:blowup-theorem} this case is excluded by
condition~\ref{cond:Euler-d2pdmu2} of Lemma~\ref{la:Euler-smooth}.

Ideally, there would be a useful interaction between the results
of~\cite{Barnes-etal:Glimm-Gowdy} and those of the present paper. It
would then be possible to conclude that the classical solutions with
finite-time breakdown can be continued as weak solutions and on the
other hand that the weak solutions which have been constructed are not
always classical.  Unfortunately the time coordinates used in the two
cases are different so that a direct comparison is not possible.

Finally, some possible extensions of the results of this paper will be
mentioned. In the case of a special relativistic fluid the analogue of
Theorem~\ref{th:blowup-theorem} holds with the same proofs. In that
case the time is the ordinary Minkowski time. Also,
Theorem~\ref{th:extension-exists} shows that under the extra condition
that $\rmd{p}/\rmd\mu$ is bounded away from $1$, a solution for which
the first derivatives of $m$ and $u$ remain bounded remains smooth,
i.e. it can be extended smoothly to a longer time interval. The
results obtained here therefore support the picture that the breakdown
of the classical solutions is due to the formation of shock waves.

\vskip 10pt\noindent
{\it Acknowledgement}
One of us (ADR) thanks Blaise Tchapnda for useful discussions. FS
would like to thank the Albert Einstein Institute for their generous
support.



\begin{thebibliography}{10}

\bibitem{Alinhac:blowup}
Alinhac, S. 1995 \emph{Blowup for nonlinear hyperbolic equations}, 
Progress in Nonlinear Differential Equations and Their Applications, vol.~17,
  Birkh\"auser, Boston.

\bibitem{Andreasson:Einstein-Vlasov-review}
Andr\'easson, H. 2005 \emph{The {E}instein-{V}lasov system/kinetic theory},
  Living Rev. Relativ. \textbf{8} no.~2,
  \url{http://www.livingreviews.org/lrr-2005-2}.

\bibitem{Barnes-etal:Glimm-Gowdy}
Barnes, A.~P., Le~Floch, P.~G., Schmidt, B.~G. and Stewart, J.~M. 2004
\emph{The
  {G}limm scheme for perfect fluids on plane-symmetric {G}owdy spacetimes},
  Class. Quantum Grav. \textbf{21} , 5043--5074.

\bibitem{Chen:conserv-SR}
Chen, J. 1997 \emph{Conservation laws for relativistic fluid 
dynamics}, Arch.
  Ration. Mech. Anal. \textbf{139} , 377--398.

\bibitem{Christodoulou}
Christodoulou, D. 2007 \emph{The formation of shocks in 3-dimensional fluids}
EMS Publishing, Zurich. 

\bibitem{Dafermos:conserv-laws}
Dafermos, C.~M. 2000 \emph{Hyperbolic conservation laws in continuum
  physics}, Grundlehren der matematischen Wissenschaften, vol. 325, Springer,
  Berlin.

\bibitem{Frid-Perepelitsa:rel-gas-dynamics}
Frid, H. and Perepelitsa, M. 2004 \emph{Spatially periodic solutions in 
relativistic
  isentropic gas dynamics}, Commun. Math. Phys. \textbf{250}  335--370.

\bibitem{Groah-Temple:EE-fluid-Glimm}
Groah, J. and Temple, B. 2004 \emph{Shock-wave solutions of the {E}instein 
equations
  with perfect fluid sources: {E}xistence and consistency by a locally inertial
  {G}limm scheme}, Mem. Amer. Math. Soc., vol 172.

\bibitem{Hormander:nonlin-hyp}
H\"ormander, L. 1997 \emph{Lectures on nonlinear hyperbolic differential
  equations}, Math\'ematiques \& Applications, vol.~26, Springer, Berlin.

\bibitem{Isenberg-Rendall:cosmo-not-CMC}
Isenberg, J. and Rendall, A. D. 1998 \emph{Cosmological spacetimes not covered
  by a constant mean curvature slicing}, Class. Quantum Grav. \textbf{15}
  , 3679--3688, gr-qc/9710053.

\bibitem{Jenssen-Sinestrari:blowup-scalar-balance-law}
Jenssen H. K. and Sinestrari, C. 1999 \emph{Blowup asymptotics for
  scalar conservation laws with a source}, Commun. PDE \textbf{24} 
  no.~11--12, 2237--2261.

\bibitem{Rendall:crushing-sings}
Rendall, A. D. 1995 \emph{Crushing singularities in spacetimes with spherical,
  plane and hyperbolic symmetry}, Class. Quantum Grav. \textbf{12},
  1517--1533.

\bibitem{Rendall:exist-nonexist-CMC}
\bysame, 1997 \emph{Existence and non-existence results for global constant 
mean
  curvature foliations}, Nonlinear Anal. \textbf{30}, 3589--3598,
  gr-qc/9608045.

\bibitem{Rendall:CMC-2dim-symmetry}
\bysame, 1997 \emph{Existence of constant mean curvature foliations in 
spacetimes
  with two-dimensional local symmetry}, Commun. Math. Phys. \textbf{189}
  , 145--164.

\bibitem{Rendall:existence-review}
\bysame,  2005 \emph{Theorems on existence and global dynamics for the 
{E}instein
  equations}, Living Rev. Relativ. \textbf{5} , no.~6,
  \url{http://www.livingreviews.org/lrr-2005-6}.

\bibitem{Rendall:book}
\bysame, 2008 \emph{Partial Differential Equations in General 
Relativity}, 
Oxford University Press, Oxford.

\bibitem{Smoller-Temple:relativistic-Euler}
Smoller, J. A. and Temple, B. 1993 \emph{Global solutions of the relativistic 
{E}uler
  equations}, Commun. Math. Phys. \textbf{156} , 65--100.

\bibitem{Taub:Rankine-Hugoniot}
Taub, A. H. (1948) \emph{Relativistic {R}ankine-{H}ugoniot equations}, 
Phys. Rev.
  , 328.

\bibitem{Yodzis-Seifert-zumHagen:naked-II}
Yodzis, P., Seifert, H.-J. and M{\"u}ller~zum Hagen, H. 1974  \emph{On the 
occurrence
  of naked singularities in general relativity. {II}}, Commun. Math. Phys.
  \textbf{37} , 29--40.

\bibitem{Yong:global-exist-balance-laws}
Yong, W.-A. 2004 \emph{Entropy and global existence for hyperbolic balance 
laws},
  Arch. Ration. Mech. Anal. \textbf{172} , 247--266.

\end{thebibliography}

\providecommand{\bysame}{\leavevmode\hbox to3em{\hrulefill}\thinspace}
\providecommand{\MR}{\relax\ifhmode\unskip\space\fi MR }
\providecommand{\MRhref}[2]{%
  \href{http://www.ams.org/mathscinet-getitem?mr=#1}{#2}
}
\providecommand{\href}[2]{#2}

\end{document}